\documentclass[pra,twocolumn,superscriptaddress,floatfix]{revtex4}

\usepackage{graphics}           
\usepackage{bm}                 
\usepackage{amsmath}            
\usepackage{amsfonts}
\usepackage{amssymb}
\usepackage{latexsym}           

\newcommand{\ket}[1]{\mbox{$|#1\rangle$}}
\newcommand{\bra}[1]{\mbox{$\langle#1|$}}
\newcommand{\braket}[2]{\mbox{$\langle#1|#2\rangle$}}
\newcommand{\identity}{\leavevmode\hbox{\small1\kern-3.2pt\normalsize1}}

\begin{document}

\title{Quantum walks on general graphs}

\author{Viv Kendon}
\email{Viv.Kendon@ic.ac.uk}
\affiliation{Optics Section, Blackett Laboratory, Imperial College,
London, SW7 2BW, United Kingdom.}

\date{\today}

\begin{abstract}
Quantum walks, both discrete (coined) and continuous time,
on a general graph of $N$ vertices with undirected edges
are reviewed in some detail.
The resource requirements for implementing a quantum walk as a
program on a quantum computer are compared and found to be very
similar for both discrete and continuous time walks.
The role of the oracle, and how it changes if more prior information
about the graph is available, is also discussed.
\end{abstract}



\maketitle


\section{Introduction}
\label{sec:intro}

One of the most important tasks on the theoretical side of quantum
computing is the creation and understanding of quantum algorithms.
The recent presentation of two quantum algorithms based on quantum
versions of random walks is particularly important in this respect,
since they provide a new type of algorithm which can show an
exponential speed up over classical algorithms, to add to those
based on the quantum Fourier transform.
Childs \textit{et al.}~\cite{childs02a} have produced a scheme for a
continuous time quantum walk
that can find its way across a particular graph exponentially faster than
any classical algorithm, while Shenvi \textit{et al.}~\cite{shenvi02a} proved
that a discrete quantum walk can reproduce the quadratically faster search times
found with Grover's algorithm for finding a marked item in an unsorted
database.
For an overview of the development of quantum walks for quantum computing,
see the recent review by Kempe \cite{kempe03a}.

The importance of classical random walk techniques for computational
physics motivates the study of quantum walk algorithms both for 
understanding in principle the power of quantum computation,
and for the possibility of a practical quantum computer that can
solve problems more efficiently.
It is important to make a fair comparison between quantum and classical
algorithms when determining whether the quantum method is really faster.
Resource counting is difficult at the best of times, and for
the quantum walk algorithms presented so far,
the presence of an oracle is an added complication.
One point that has not been made explicit by most authors to date
is that if, as is usually the case, the outcome of the quantum walk will be
obtained by measuring the position,
the quantum walk algorithm should have the position space encoded
into a qubit register (or equivalent \cite{blume02a}).
This is simply because for a classical random walk algorithm,
the location of the particle can be encoded in a binary string, so we will
penalize ourselves by requiring exponentially more resources if we adopt
unary encoding for our quantum walk.
Childs \textit{et al.}~\cite{childs02a} explicitly perform this encoding,
but without remarking on this point, since their primary motivation is
to show that they can implement a \textit{continuous} time Hamiltonian
evolution efficiently on a \textit{discrete} quantum circuit.
On the other hand, this places the proposed physical implementations
of a quantum walk \cite{travaglione01a,sanders02a,dur02a} firmly in
the realm of physics, as all of them have the position space of the walk
set up in such a way that it cannot be measured as a binary encoded
bit string.  None of them claim they are trying to perform algorithms
with these walks, and the experiments are of considerable interest in
their own right as examples of coherent control in physical systems.
For a discussion of the fundamental characteristics of quantum walks in
physical systems, see Ref.~\cite{sanders03a}.

There are important open questions surrounding the relationship
between the two main formulations of quantum walks in the literature,
discrete time \cite{aharonov92a,watrous98a,aharonov00a,ambainis01a},
and continuous time \cite{farhi98a}.  
Classically, the discrete and continuous versions of random walks can be
related in a straightforward manner by taking the limit of the
discrete walk as the size of the time step goes to zero.
In the quantum case, the way the systems are commonly formulated,
the discrete time and continuous time versions have different sized
Hilbert spaces so one cannot be the limit of the other without showing that
the larger system is actually restricted to a subspace the same size as the
smaller under suitable conditions.  Also open, but generally expected to be
true, is whether discrete and continuous time quantum walks have the
same computational power.  They give broadly similar results for 
algorithmic properties such as mixing times and hitting times
in all cases where both forms have been applied to the same problem
though their detailed dynamics are different.

The main aim of this paper is to describe a quantum walk
process for both discrete (coined) and continuous
time quantum walks on a very general graph side by side, so they
can easily be compared and contrasted.
The general case of the discrete time quantum walk has only appeared
in the literature in Watrous \cite{watrous98a}, in the form of
a unitary process on a directed graph derived from the undirected graph
under study.
This formulation is exactly equivalent to using a coin combined with
the original (undirected) graph, but physicists
seem to prefer the coined version, so it
is useful to write out the coined quantum walk in the general case.
We then observe that when a continuous time quantum walk is
implemented as a quantum computer program, the structure of that
program turns out to be very similar to the discrete time walk
on the same graph.

\section{Classical random walk on general graph}
\label{sec:classgen}

We start by defining the type of graph we will consider here.
A graph $G(V,E)$ has a set of $N$ vertices $v\in V$ and a set of
edges $e\in E$ joining pairs of vertices.
The edges are \textit{undirected}, that is,
$e_{xy} \equiv e_{yx}$ for the edge joining vertices $v_x$ and $v_y$.
We assume there is at most one edge between any pair of vertices.
The maximum number of edges that can meet at each vertex
-- the degree of the graph -- is thus $(N-1)$, and
the maximum number of edges we could have in total is $N(N-1)/2$, one between
every possible pair of vertices.  The degree of a vertex is 
the number of edges that meet at it, and can be zero if there are no
edges joining that vertex to any other.

The description of the graph is conveniently expressed as the adjacency
matrix $\mathbf{A}$, which is of dimension $N\times N$, symmetric,
with element $A_{xy} = 1$ iff $e_{xy} \in E$ and zero otherwise.  
Since there are $N(N-1)/2$ possible edges, $\mathbf{A}$ actually
contains two copies of this information, but that duplication allows
us to use a matrix rather than a triangular table, and avoids having to first
determine whether $x>y$ or vice versa when reading out values of $A_{xy}$.
The ordered way in which information is stored in $\mathbf{A}$
is an example of a hash table that allows any bit within it
to be accessed with a single query specifying the values of $x$ and $y$.
The full information describing a single
edge needs to be given not by a single bit, but by the set $(x,y,A_{xy})$
that identifies the edge by its vertex indices.
With no other prior information about the graph, the adjacency matrix
is the most efficient way to store and access the description.

Already our general graph has several conditions attached to it,
undirected edges and only one edge between any pair of vertices.
Graph theory is such a broad field that variants suggest themselves
almost anywhere you look.  Directed edges (that can be traversed in
only one direction) give a completely different character to the
graph, like the difference between reversible and irreversible processes.
They are beyond the scope of this paper (for reasons which will become
clear later), the reader can refer, for example, to
Severini \cite{severini02a,severini03a} where a (different formulation of)
discrete quantum walks on directed graphs is discussed.

On the other hand, classically, multiple edges between a pair of
vertices are equivalent to a single edge with 
a weight attached that modifies the probability of choosing that
path to the next vertex.
The generalization to weighted edges is straightforward, just replace
the bits in the adjacency matrix by appropriately normalized weights.
In the quantum case, different paths joining the same point can lead
to interference effects, so multiple edges between two vertices are
not equivalent to a single edge with a weight.
We can also have complex weights for quantum walks,
since they can be used to modify the amplitude of the wave function,
rather than the probability of choosing that edge.
An alternative approach to multiple edges is to reduce it to
the case of single edges by adding a vertex to each extra edge.
This increases the size of the graph ($N$) but keeps the degree of
each vertex bounded (by $N-1$).
However, this reduction is not equivalent for all properties of interest
on a graph, in particular for dynamical or flow properties, where the extra
vertices and edges will slow down the progress of the random walk.
We will not treat graphs with weights or multiple edges in this paper,
though we note that this leaves interesting open questions about which
types of graphs and properties are most amenable to analysis using quantum
versions of random walks.

\subsection{Discrete time classical walk}
\label{ssec:classdis}

A classical random walk on a graph $G(V,E)$ can be performed as follows.
Describe the location of the walk by an index $x$ with
$0\le x\le (N-1)$ running over the vertices in the graph.
At each time step, a random number $y$ between $0 \le y \le (N-1)$
is generated, and the value of $A_{xy}$ examined to find out if
edge $e_{xy}$ exists connecting the current location $v_x$ to $v_y$.
If the vertices are connected, the location index is updated to be $y$,
(possibly only with some probability $\alpha$,
to tune the properties of the walk),
and if not, the walk remains where it is at $x$.
This procedure is repeated for as many steps as required.
Note that we haven't bothered to exclude the case $y=x$.  For large $N$,
it isn't worth the extra computational time to check this condition for
every new $y$ generated.

The probability $P(x,t)$ of finding
the walk at location $x$ at time $t$ is given by the difference
equation representing one iteration of the walk,
\begin{eqnarray}
P(x,t+1) &=&P(x,t) +\nonumber\\
&&\frac{\alpha}{N}\sum_y\left\{ A_{xy}P(y,t) - A_{yx}P(x,t)\right\}.
\label{eq:cdis}
\end{eqnarray}
Since $\sum_y A_{yx} = d_x$ is the number of edges connected to
vertex $v_x$, there are various useful ways to rewrite this equation,
especially if the values of $d_x$ are known in advance.

\subsection{Continuous time classical walk}
\label{ssec:classcon}

A continuous time classical random walk on this graph can be
implemented in a similar way.  We now have a hopping rate
$\gamma$ per unit time to connected vertices,
and the probability distribution can be written
\begin{equation}
 \frac{d}{dt}P(x,t) =\gamma\sum_y\left\{ A_{xy}P(y,t) - A_{yx}P(x,t)\right\}.
\label{eq:ccon}
\end{equation}
As written, this is clearly the continuous time limit of the discrete
time version given in Eq.~(\ref{eq:cdis}), with time rescaled to
$t/N$, and $\gamma = \alpha$.
This formulation of a random walk is more useful where we have
some type of analytic solution to Eq.~(\ref{eq:ccon}),
such that the implementation has advantages over the discrete time random walk,
or where it more closely matches a physical process we are trying to model.

\subsection{Resources required for implementation}
\label{ssec:cres}

To quantify the resources required for implementing a
random walk as an algorithm on a classical computer,
observe that we can code the location index $x$ in
a binary string of size $n=\lceil \log N\rceil$ bits,
exponentially smaller than the description of the graph ($\mathbf{A}$).
We can assign the task of storing $\mathbf{A}$ to an oracle,
and obtain the value of $A_{xy}$ by querying the oracle with the pair of
numbers $(x,y)$ at each step.  This means the resources for performing
the algorithm can be exponentially smaller than the description of the
graph.  By assigning the description of the graph to an oracle, we are
separating out the resources required to describe the problem from the
resources required to perform the algorithm.  
Specifying a problem or an answer to a problem can be highly
non-trivial if it takes exponential resources to describe.  A good example
is solutions to Pell's equation, for which a quantum algorithm has been
given by Hallgren \cite{hallgren02a}.

This distinction is not merely formal.  If we want to sample the
properties of a random graph, we do not necessarily need to generate
the whole description of the graph, we can just generate values of
$A_{xy}$ on the fly as we sample them, allowing us to study the
properties of structures larger than our computational resources
can store.
Also note that if we want to keep the algorithm small, i.e.,
$O(\text{poly}(n))$, we cannot afford to increase the efficiency 
by remembering the values of $(x,y)$ which we have already queried,
since there are exponentially many, $O(N\simeq 2^n)$ of them even for one vertex.
It follows that it can take an exponential number of steps to make useful
progress on the graph, since the probability of making a move is
$\alpha d_x/N$.
If $d_x$ is $O(1)$, the chance of making a move is exponentially small.

In the case of the continuous time random walk, the resource requirements
are similar except we
also have to account for the resources required to implement a sufficiently
accurate numerical approximation of the continuous time dynamics,
possibly over exponentially long times, $t\sim O(N)$, or $t\sim O(N^2)$.

\subsection{More information about the graph}
\label{ssec:moreinfo}

The general graph we have chosen to define can thus be analyzed using
a random walk algorithm with small requirements for memory, $O(\text{poly}(n))$,
but large requirements for time, $t \sim O(\text{poly}(N))$.
For many algorithmic and computational complexity purposes it
makes sense to identify a class of problems for which the time
requirements can also be small.  This is the case for all the examples
in the literature on which quantum walk dynamics have been analyzed.
It means we either need to restrict ourselves to problems where
$N$ is not ``exponentially large'', or else we need to know more about
the graph, so that we can guarantee to move to a new location in
$O(\text{poly}(n))$ steps, rather than $O(N)$.
On the general graph we have no way of knowing ahead of time whether
any pair of vertices is connected by an edge or not.  But if we are given,
say, a regular two-dimensional lattice with some of the edges
missing, we can draw the random location index $y$ from a much more
restricted set (nearest neighbors, assuming they are labeled in
a predictable way) and have a much higher success rate in finding the
next move for the random walk.  If we are doing a simple random walk on
a line, cycle or hypercube,
then we know exactly which edges are connected to any vertex,
and the random choice is simply to pick which one to take.

This reduction happens in general because
we can encode the description of a more regular graph in a more
efficient way than the adjacency matrix.
If the description is exponentially smaller than the number of vertices
$N$, then the description can be incorporated into the algorithm at no
significant extra cost, and the oracle has no role to play in the analysis.
Most of the quantum walks in the literature to date are on sufficiently
regular structures that fall into this category.
The main exception is the ``glued trees'' graph of Ref.~\cite{childs02a},
which has half the edges in a random join down the middle of the
graph.  This still allows a quadratic reduction in the description
of the graph, since the degree of the graph is fixed and $O(1)$,
so there are only $O(N)$ edges to describe.
However, the oracle is necessary here 
if we wish to use resources only $O(\log N)$ for the algorithm itself. 

Though the description of the random walk process remains qualitatively the same
in words, if the graph is now specified in some other, more efficient
way than the adjacency matrix, Eqs.~(\ref{eq:cdis}) and (\ref{eq:ccon})
are no longer valid and must be replaced by versions using the 
appropriate graph specification reflecting the higher probability of making
a move.  It may be sufficient to adjust the value of $\alpha$ or $\gamma$,
but often the equation for the dynamics of the random walk
ends up looking rather different, once $A_{xy}$ has been replaced by
a formula for generating the neighboring vertex indices, for example.

\section{Quantum walks on general graphs}
\label{sec:qrwgen}

A quantum version of a random walk process must at least have the
classical random walk as its classical limit.  But this does not
uniquely specify the approach to take.
For a quantum evolution we require the dynamics to be unitary, and hence
reversible, so the random element of the algorithm has to be catered for
in other ways, for example, in the random result from a measurement with
several possible outcomes.  This gives us one route to reproduce a
classical random walk from the quantum dynamics, by applying a measurement
at every step of the walk.
Here we will first consider discrete time versions of quantum walks,
even though historically for a general graph such as we are considering,
the continuous time quantum walk was presented first, by Farhi and Gutmann
\cite{farhi98a}.  There are some subtleties to do with the 
continuous time case that will be best appreciated after
the discrete time version has been described.

\subsection{Discrete quantum walks on general graphs}
\label{ssec:qdis}

Let us first consider the most na\"ive way in which the classical discrete
random walk on a general graph could be quantized.
The location of the walk is represented by a state in a Hilbert space
of size $N$, and may thus be in a superposition of positions.
Intuitively, what we would like to do at each step of the walk
is to distribute the wave function equally along all the
connected edges from each vertex, instead of just picking one at random.
However, the adjacency matrix is not unitary in general, so we cannot simply
apply it as an operator to the particle state in a quantum version
of Eq.~(\ref{eq:cdis}).
We could try to make it unitary by introducing
phases so the non-zero entries are $e^{i\phi_{xy}}$
rather than unity, giving a quantum ``adjacency matrix'' $\mathbf{A^{(Q)}}$ and
\begin{eqnarray}
\ket{x,t+1} &=& \ket{x,t} + \frac{\alpha}{N}\sum_{y}\left\{
	A^{(Q)}_{xy}\ket{y,t} - A^{(Q)}_{yx}\ket{x,t}\right\}\nonumber\\
&=& \sum_{y}\left\{\frac{\alpha}{N}A^{(Q)}_{xy}
	+ \left(1 - \frac{\alpha d_x}{N}\right)\delta_{xy}\right\}\ket{y,t}
\end{eqnarray}
for the evolution.  This only produces a unitary operator under
very restricted conditions, see Meyer \cite{meyer96a} who shows that this 
does not work for quantum walks on Euclidean lattices,
and Severini \cite{severini02a,severini03a}
for the more general case of graphs with directed edges.

Instead, we need to model our quantum walk more closely
on the algorithm for a classical random walk.
This gives us a unitary process for any undirected graph.
In the classical discrete random walk, we query the oracle with the
pair of vertices $(x,y)$ to obtain the value of $A_{xy}$.
For $A_{xy}=1$, the value of $x$ is replaced with $y$ and a new $y$
generated for the next step.  
A quantum version of this procedure,
with a superposition of positions instead of a single $\ket{x,y}$,
needs to have the oracle receive and return the
whole pair $\ket{x,y}$ along with a further qubit to hold the
value of $A_{xy}$.  This is to ensure the operation is unitary,
and also because we can't make and keep a copy of the state
while passing $\ket{x,y}$ to the oracle.
We can then perform a conditional swap
operation depending on the value of $A_{xy}$, to update the position of
the walk.  The quantum oracle $\mathcal{O}$ is defined by
\begin{equation}
\mathcal{O}\ket{x,y,0} = \ket{x,y,A_{xy}},
\label{eq:qdiso}
\end{equation}
and the conditional swap $\mathcal{S}$ is defined by
\begin{equation}
\mathcal{S}\ket{x,y,A_{xy}} = \left\{ \begin{array}{lr}
				      	\ket{y,x,A_{xy}} \/\/ 	& A_{xy} = 1 \\
					\ket{x,y,A_{xy}}	& A_{xy} = 0 \\
				\end{array}
				\right..
\end{equation}
We aren't finished yet.  We need to reset the value of the qubit holding
$A_{xy}$ to zero, and generate a new value of $y$ ready for the next
oracle query.  The first task can be done with a second call to the oracle,
given the oracle is unitary and symmetric, and therefore its own inverse,
\begin{equation}
\mathcal{O}\ket{x,y,A_{xy}} = \ket{x,y,0},
\end{equation}
and since for the graphs we are considering, $A_{xy}=A_{yx}$ it doesn't
matter that some of the $x,y$ pairs have been swapped.
The value of $y$ can be updated using any reasonable choice of unitary 
operation as a quantum ``coin toss'' $\mathcal{C}$, that may be
conditioned on the value of $x$ if we want.
Actually, the usual order is to toss the coin first,
then apply the oracle-swap operations, so one
step of the quantum walk is written
\begin{equation}
\ket{\psi(t+1)} = \mathcal{OSOC}\ket{\psi(t)},
\end{equation}
and $t$ steps can be written
\begin{equation}
\ket{\psi(t)} = (\mathcal{OSOC})^t\ket{\psi(0)},
\label{eq:qdissol}
\end{equation}
where $\ket{\psi(0)}$ can be the all zero state for walks starting at
vertex zero.
The only factor missing compared to the classical algorithm
is $\alpha$, determining whether we actually
take the step (when allowed) or not.  The quantum alternative is to tune the
properties of the walk by adjusting the coin toss operator $\mathcal{C}$,
if we want to keep the whole process unitary.
This gives us more ``knobs to twiddle'' than we defined in the classical case,
but in practice there are also many more ways to tune a classical random walk
process than we have mentioned here.  We could, for example, have $\alpha$
be vertex dependent, $\alpha(x)$, or even edge dependent, $\alpha(x,y)$,
which is equivalent to weighted edges.  Similarly, in the continuous time
walk, $\gamma$ can be vertex or edge dependent.

\subsection{Resources required for discrete walk}
\label{ssec:dqres}

It is instructive to compare and contrast the resources needed for this
quantum walk with the classical case.  We have the same number
of qubits as classical bits, i.e., $2n+1$, with $n=\lceil\log N\rceil$.
For the coin toss $\mathcal{C}$ on the $y$ register,
we should restrict ourselves to unitary operators that can be
applied efficiently using $O(\text{poly}(n))$ gates, or else
we will have an exponentially inefficient algorithm.
The simplest example would be to apply a Hadamard
to each qubit in $\ket{y}$ individually, but the optimal choice
will depend on the graph and problem to be solved \cite{tregenna03a}.
The controlled swap $\mathcal{S}$ can be done efficiently as a sequence of
Fredkin gates applied to each corresponding pair of qubits.

We need a quantum oracle rather than a classical oracle, which is capable
of processing and returning quantum superposition states.  Given this
quantum oracle, we need just two calls to it per step, whereas the
classical algorithm has to query the classical oracle $N/d_x$
times on average for each actual move made.
This certainly makes the quantum algorithm look more powerful for $d_x<N/2$,
since it can move up to $m/2$ steps for $m$ oracle calls, compared to an
average of $m d_x/N$ steps for $m$ oracle calls for the classical walk.
But the quantum oracle is not the same as the classical oracle,
which only returns one classical bit per query,
so we need to be careful before drawing conclusions.

If we insisted that the quantum algorithm must use the classical oracle,
then it will need to retrieve the whole adjacency matrix in order to perform
unitary operations that correspond to each possible pair of values of
$x$ and $y$ that might be in superposition in the quantum state \ket{\psi(t)}
of the walk.  On the other hand, the classical walk could use the quantum
oracle with the help of a qubit state preparation device,
by querying it with a single pair of values $\ket{x,y,0}$ and
receiving $\ket{x,y,A_{xy}}$ in return, which could then be measured to obtain
$A_{xy}$ with certainty.

To illustrate further that the oracle--algorithm separation is fairly arbitrary
in this situation, we can redefine the oracle to be the whole conditional swap
operation $\mathcal{OSO}$.  Classically this just means we give the pair
$(x,y)$ to the oracle, and receive in return either $(x,y)$ if
there is no edge connecting those two vertices, or $(y,x)$
if there is a connecting edge.
Since the numbers are classical bits, we
could inspect their values and detect whether the swap took place,
but we don't need to know, we just toss a new value for $y$ and continue;
we no longer need the single bit for the value of $A_{xy}$.
The quantum version of this oracle is
\begin{equation}
\mathcal{OSO} = \sum_{xy|e_{xy}\in E}   \ket{y,x}\bra{x,y} +
	        \sum_{xy|e_{xy}\notin E}\ket{x,y}\bra{x,y},
\label{eq:gstateo}
\end{equation}
where we also need one less qubit, since the value of $A_{xy}$ is not
returned.

On a more basic level than resource counting, we have not shown that
this definition of a discrete quantum walk does anything usefully different
from a classical random walk in the general case.  However, 
the specific cases of coined quantum walks in the literature are
clearly derivable from it by encoding the simpler graph more
efficiently.
For example, the walk on a hypercube is obtained by
noting that there are exactly $n$ choices of connected vertices rather
than a possible $(N-1)$, and shrinking the size of the $\ket{y}$ qubit
register accordingly to $\lceil\log n\rceil$.
The oracle operation is redundant (unless the
labeling of the vertices has been randomized) since the value of
$y$ is used to specify which bit of the value of $x$ to update to
obtain the new vertex index.
Useful classical random walk algorithms are also usually highly tuned
to suit the problem, so we shouldn't necessarily expect that a
proof of quantum superiority over classical would be forthcoming in
the general case, as it will depend on the particular choice of coin
toss operation \cite{tregenna03a} and the particular property of
the graph we are trying to determine.
It has already been shown that mixing times on a hypercube are not faster
than classical \cite{moore01a}, whereas the hitting times are,
at least for the equivalent classical random walk process \cite{kempe02a}.

\subsection{Continuous time quantum walks on general graphs}
\label{sec:qrwgencon}

The continuous time quantum walk on a general graph,
as presented by Farhi and Gutmann \cite{farhi98a},
simply uses the adjacency matrix
$\mathbf{A}$, which is Hermitian for an undirected graph,
to form the Hamiltonian for the evolution of the quantum state,
\begin{equation}
i\frac{d}{dt}\braket{x}{\psi(t)} =
	\sum_y\bra{x}\mathbf{H}\ket{y}\braket{y}{\psi(t)},
\label{eq:qcon}
\end{equation}
with $\mathbf{H}=\gamma\mathbf{A}$, and solution 
\begin{equation}
\ket{\psi(t)} = 
e^{-i\gamma\mathbf{A}t}\ket{\psi(0)}.
\label{eq:qconsol}
\end{equation}
Comparing this with Eq.~(\ref{eq:ccon}) shows that the
second term, necessary for conservation of probability,
is missing.
Since we only need a Hermitian operator here, and since 
$A_{xy}=A_{yx}$ guarantees this, we are free to examine this
quantum evolution as well as that obtained in more direct analogy to
Eq.~(\ref{eq:ccon}),
\begin{eqnarray}
i\frac{d}{dt}\braket{x}{\psi(t)} =
	\gamma\sum_y&&\left\{\bra{x}\mathbf{A}\ket{y}\braket{y}{\psi(t)} 
			\right.\nonumber\\
		    &&-\left. \bra{y}\mathbf{A}\ket{x}\braket{x}{\psi(t)}\right\}.
\label{eq:qcon2}
\end{eqnarray}
For graphs where all the vertices are of the same degree $d$,
the second term only introduces an irrelevant global phase,
therefore it makes no difference to observable quantities if it is omitted
\cite{ahmadi02a}.
The Hamiltonian in Eq.~(\ref{eq:qcon2}) becomes
$\mathbf{H}=\gamma(\mathbf{A} - d\identity)$,
and the solution to this can be written
\begin{equation}
\ket{\psi(t)}=e^{-i\gamma(\mathbf{A}-d\identity)t}\ket{\psi(0)}.
\end{equation}
Since $\mathbf{A}$ commutes with the identity, the two
terms in the exponential can be factored, giving 
\begin{equation}\ket{\psi(t)}=
e^{-i\gamma\mathbf{A}t}e^{i\gamma d\identity t}\ket{\psi(0)},
\end{equation}
which is the same as Eq.~(\ref{eq:qconsol}) apart from the phase
$e^{i\gamma d\identity t}$.
For graphs of general degree, however, the dynamics with the
second term included will have a different evolution \cite{childs03a}.
So far, only graphs of fixed degree have been studied in any detail in
the literature so the difference between the two versions has not been
explored.

Equation (\ref{eq:qconsol}) looks remarkably similar to the discrete
case, Eq.~(\ref{eq:qdissol}), both can be written in the form
\begin{equation}
\ket{\psi(t)}=(\mathcal{U})^t\ket{\psi(0)},
\end{equation}
with unitary operator $\mathcal{U}=\mathcal{OSOC}$ for the discrete time walk
and $\mathcal{U}=e^{-i\gamma\mathbf{A}}$ for the continuous time walk.
But, unlike the classical case, where Eq.~(\ref{eq:ccon}) is the limit 
of Eq.~(\ref{eq:cdis}) as the time step goes to zero, in the quantum case
the similarity is deceptive, the discrete and continuous walks have
Hilbert spaces of different sizes, $H_N$ in the continuous case and
$H_N\otimes H_N \otimes H_2$ in the discrete case.  
Moreover, though the $H_2$ component returns to $\ket{0}$ at
the end of each step and thus factors out, the discrete walk evolution does
not remain in an $H_N$ subspace (apart from trivial special cases),
so this is not a simple route to showing how the two versions are related.

\subsection{Resources required for continuous walk}
\label{ssec:cqres}

For algorithmic purposes we need to encode the walk into $n$-qubit registers,
and evaluate Eq.~(\ref{eq:qconsol}) for our chosen values of $t$ and
$\ket{\psi(0)}$.  Childs \textit{et al.}~\cite{childs02a},
in their quantum walk algorithm, showed how to accomplish this for a graph
of degree O(1), where their oracle held a more efficient description of the
graph than the adjacency matrix.
Here we will describe their method as applied to a general graph, using
the same oracle $\mathcal{O}$ as in the discrete time quantum walk 
defined by Eq.~(\ref{eq:qdiso}).

We take a qubit register of size $n$ to
hold the vertex index indicating the current location of the walk,
and two ancillary registers, one also of
size $n$ qubits and the other a single qubit.
The Hilbert space is thus of size $H_N\otimes H_N\otimes H_2$, the
same size as for the discrete time quantum walk in Sec.~\ref{ssec:qdis}.
First we will describe what goes wrong if we try to avoid
introducing a ``consistent coloring'' of the edges.
Define the set of unitary operators $\mathcal{V}_y$
that write the value of $y$ into the second register,
\begin{equation}
\mathcal{V}_y\ket{x,z,b} = \ket{x,z\oplus y,b},
\end{equation}
where $\oplus$ specifies bitwise addition, and we will take the initial
values of $z$ and $b$ to be zero.
Note that $\mathcal{V}_y$ is its own inverse.
After applying $\mathcal{V}_y$, we can apply the oracle $\mathcal{O}$
to place the value of $A_{xy}$ in the last qubit.
Since the actual input can be a superposition of any of the $N$ vertex indices
$x$, the operation $\mathcal{O}\mathcal{V}_y$ can produce a superposition of
all possible $A_{xy}$ for a given $y$.
Next we apply the Hermitian evolution that does a conditional swap on the first
two registers, and keeps only the terms in the graph with $A_{xy}=1$,
as indicated by the final qubit,
\begin{eqnarray}
\mathbf{T}\ket{x,y,1} &=& \ket{y,x,1} \nonumber\\
\mathbf{T}\ket{x,y,0} &=& 0.
\end{eqnarray}
This can also conveniently be written
\begin{equation}
\mathbf{T}\ket{x,y,A_{xy}} = A_{xy}\ket{y,x,A_{xy}}.
\end{equation}
Childs \textit{et al.}~\cite{childs02a} show that $e^{-i\mathbf{T}t}$
can be evaluated efficiently using $O(\text{poly}(n))$ gates,
and give a circuit diagram for this.  We will not repeat these steps here.
Next we must reset the extra registers to zero, in order to keep the
evolution of the walk in the subspace of the first $n$ qubits.
Removing the value of $A_{xy}$ from the last qubit is accomplished
with another application of the oracle $\mathcal{O}$, but, because
the first two registers have been swapped, we now need to apply
$\mathcal{V}_x$ to get what we want
\begin{equation}
\mathcal{V}_x\ket{y,x,0} = \ket{y,0,0}.
\end{equation}
However, since $\mathcal{V}_x \neq \mathcal{V}_y$,
this won't give us a Hermitian operator overall.
This is where the consistent coloring comes in.
The edges are given labels (colors) such that all vertices have
at most one edge with each color.  Then the operators are
labeled by color instead of by the vertex they connect to, and
the quantity that is added to the second register is $y_c(x)$, the index of
the vertex connected to $v_x$ by color $c$, 
\begin{equation}
\mathcal{V}_c\ket{x,z,b} = \ket{x,z\oplus y_c(x),b}
\label{eq:Vcdef}
\end{equation}
Again, $\mathcal{V}_c$ is its own inverse, but now it also provides
the required operation when applied after $\mathbf{T}$ to reset
the auxiliary registers
\begin{equation}
\mathcal{V}_c\ket{y,x,0} = \ket{y,x\oplus y_c(y),0} = \ket{y,0,0}.
\end{equation}
So
\begin{equation}
\mathbf{H} = \gamma\sum_{c=0}^{N-1}\mathcal{V}_c\mathcal{O}\mathbf{T}\mathcal{O}\mathcal{V}_c
\label{eq:qconh}
\end{equation}
is a Hermitian operator that enacts the dynamics in Eq.~(\ref{eq:qconsol}).
This can easily be checked by calculating $\bra{y,0,0}\mathbf{H}\ket{x,0,0}$
and showing that it equals $\gamma A_{xy}$.
Also, since $\mathbf{H}\ket{x,0,0} = \sum_c A_{x,y_c(x)}\ket{y_c(x),0,0}$,
provided the walk starts in the subspace spanned by the first set of $n$
qubits, it remains in this subspace at the end of the walk.

A consistent coloring is always possible using one
more color than the maximum degree of the graph \cite{vizing64a}.
Here, since we only know the maximum degree
of the graph is bounded by $(N-1)$, we have to assume a coloring using
$N$ colors.  This at least makes the coloring trivial to implement,
since a coloring that works for the complete graph (all vertices joined
to all others) will work for any other graph.
Such a coloring can be generated by using $c=(x+y)$ (mod $N$)
to create the color label $c$ for edge $e_{xy}$.

To simulate a Hamiltonian given as a sum of terms as
in Eq.~(\ref{eq:qconh}), we make use of the Lie product formula,
\begin{eqnarray}
&&e^{-i(H_1+H_2+\dots+H_N)t} \nonumber\\
&&= \lim_{j\rightarrow\infty}\left(
e^{-iH_1t/j}e^{-iH_2t/j}\dots e^{-iH_Nt/j}\right)^j,
\end{eqnarray}
only we use a finite value of $j$ chosen to give the desired accuracy.
Lloyd \cite{lloyd96a} showed this is possible using
$j\sim O(\text{poly}(N))$, where $N$ is the number of terms in the sum.
We will thus have a total of $Nj \sim O(\text{poly}(N))$ operators of the form
$\mathcal{V}_c\mathcal{O}e^{-i\gamma T t/j}\mathcal{OV}_c$ to perform.

For the specific case treated by Childs \textit{et al.}, 
this scheme provides an efficient algorithm in the sense
that it requires only $O$(poly($\log N$)) gates to implement
because they know the degree of the graph is $O(1)$
[in fact $d=3$ apart from two special vertices with $d=2$],
so the number of terms in the sum over colors in
the Hamiltonian only needs to be $O(1)$.
However, in the general case we are considering, where the degree
of the graph is bounded only by $N-1$, we have $N$ terms in our
sum over colors, so we end up with $O$(poly($N$)) gates rather
than $O$(poly($\log N$)).
This should be seen as a consequence of the general nature of the graph
we defined, see Sec.~\ref{ssec:moreinfo}, rather than a failure to
find an efficient algorithm.  In the discrete time quantum walk, we
may expect to need $O(\text{poly}(N))$ time steps, and thus
$O$(poly($N$)) quantum gates too.

The correspondence between the quantum computer programs for the discrete
and continuous time walks goes deeper than this.  In order to provide
a finite precision implementation of the continuous time walk on a 
quantum computer, we have broken down the
continuous time Hamiltonian evolution into discrete steps thus,
\begin{equation}
\ket{\psi(t)} = \prod_{c=0}^{N-1} \left(\mathcal{V}_c\mathcal{O}
		e^{-i\gamma T t/j}\mathcal{OV}_c\right)^j\ket{\psi(0)}.
\label{eq:qconprod}
\end{equation}
If we insert $\identity=\mathcal{V}_{N-1}\mathcal{V}_{N-1}$ at the start
and rewrite slightly as
\begin{equation}
\ket{\psi(t)} = \mathcal{V}_{N-1}\prod_{c=0}^{N-1} \left(\mathcal{O}
                e^{-i\gamma T t/j}\mathcal{OV}_c\mathcal{V}_{c-1}\right)^j
		\mathcal{V}_{N-1}\ket{\psi(0)},
\label{eq:qconprod2}
\end{equation}
where we take $c-1$ (mod $N$) so when $c=0$ we have $c-1=N-1$,
then the correspondence with the discrete walk,
$\ket{\psi(t)}=(\mathcal{OSOC})^t\ket{\psi(0)}$, Eq.~(\ref{eq:qdissol})
becomes clear.
A pair of $\mathcal{V}_c$ operators take the place of the coin toss
$\mathcal{C}$, and the evolution $e^{-i\gamma\mathbf{T}t/j}$ replaces
the controlled swap $\mathcal{S}$.  These operators even perform the same
function as their discrete counterparts: choosing a new value of $y$, and
swapping the $x$ and $y$ values when $A_{xy}=1$, respectively.
The pair $\mathcal{V}_c\mathcal{V}_{c-1}$ varies from step to step as $c$
varies, so this evolution is using a variable coin toss operator as well
as a different sort of swap operation, but the structure is essentially
the same.

As already noted, there are $Nj$ steps to this discretization of the
continuous time quantum walk.  Each unitary operator besides the oracle
can be performed efficiently, using $O$(poly($\log N$)) gates, so each step
uses roughly the same number of qubits and gates as a
single step of the discrete time walk.
Moreover, the discrete and continuous versions of the quantum walk are
likely to be run for roughly the same number of steps.
There is thus no essential difference between the algorithms in terms of the
number of oracle calls or the number of quantum gates.

\vspace{2em}

\section{Summary and Discussion}
\label{sec:Conc}

Quantum walk algorithms give us a new type of algorithm
to run on our quantum computers (when they are built), but
physical quantum walk processes don't give us a new type of
quantum computer architecture unless the walk takes place
in an efficiently encoded Hilbert space rather than a physical
position space.

The discrete and the continuous time quantum walk algorithms
on a general graph have similar structures when implemented on a quantum
computer, with similar efficiency and resource requirements.
This does not prove their equivalence in any algorithmic sense, but it
does provide supportive evidence for the two formulations having the same
computational power.

All the quantum walk algorithms discussed here depend on the graph
being undirected, because they rely on the adjacency matrix being symmetric.
So there are parallels with algorithms based on the quantum Fourier transform,
which can solve the hidden subgroup problem for Abelian groups and
some extensions, but not in general for non-Abelian groups.

The question of how the discrete time quantum walk relates to the
continuous time quantum walk remains a key open question for the
theoretical development of quantum walk algorithms.
A more practical challenge is to look for quantum measurements
that extract the kind of information typically gathered from a classical
random walk process.  These are often average or aggregate quantities
that may conceivably be accessible with a single measurement,
even though direct measurement of the
quantum state of the quantum algorithm would yield only one
value for the position of the walking particle.

\begin{acknowledgments}

I thank many people for useful and stimulating discussions of quantum walks,
especially Peter H\o{}yer, Richard Cleve and Barry Sanders for pointing
out errors and omissions in earlier drafts of the paper, and
Dorit Aharonov,
Ed Farhi,
Will Flanagan,
Julia Kempe,
Peter Knight,
Rik Maile,
Cris Moore,
Eugenio Roldan,
Alex Russell,
John Sipe,
Ben Tregenna,
and
John Watrous.
This work was funded by the UK Engineering and Physical Sciences
Research Council grant number GR/N2507701.

\end{acknowledgments}




\bibliography{qrw,qit}



\end{document}